\documentclass[proceedings, preprint]{rmaa}

\usepackage{paralist}

\usepackage{psfrag,color}
\usepackage{html}

\def\rsun{R$_{\odot}$}

\SetYear{2009}
\SetConfTitle{The interferometric view on Hot stars}

\title{Hot stars and interferometry}

\author{
  F. Millour\altaffilmark{1}
}

\altaffiltext{1}{Max-Planck Institute for Radioastronomy, Auf dem
  H\"ugel, 69, 53121, Bonn, Germany.}

\shortauthor{Millour}
\shorttitle{Hot stars and interferometry}

\listofauthors{F. Millour}
\indexauthor{F. Millour}

\abstract{
  What is long-baseline optical/IR stellar interferometry? A few years
  ago, many astronomers might not have been able to answer that
  question properly. This is today hopefully not the case anymore,
  because mainstream facilities, such as the VLTI, the Keck-I or the
  CHARA array, offer now this delicate technique to an astronomer who
  wants to observe his favourite object at the highest angular
  resolution available. The large teaching effort on what is
  interferometry and for what purpose it can be used, together with
  weak, but already convincing imaging capabilities, make the
  technique reaching a ``mature'' state. I will not discuss here the
  details of the technique, as already many booklets are now published
  on the subject, but rather describe what makes long-baseline stellar
  interferometry attractive for the field of hot star astrophysics.
}

\addkeyword{Stars: general}
\addkeyword{}

\begin{document}
\maketitle

\section{What do we measure with an interferometer?}
\label{sec:whatMeasure}

Long-baseline optical/IR interferometry measurements are often
exclusively considered to be the so-called ``visibility''
measurement. We need to come back to what this ``visibility'' is
linked to, to understand that it is not the only measurement made
available by interferometers. The visibility is related to the
so-called spatial coherence function of light, well described in
e.g. \citet{2007NewAR..51..565H, 2008NewAR..52..177M}. This coherence
function is proportional to the Fourier-transform of the
space-distribution of light from the observed object. 

What is often called ``visibility'' is in fact the normalized
amplitude of this Fourier-transform. However, a proper definition of
``visibility'' contains both the amplitude and the phase of this
coherence function, and we will see that the phase contains also
interesting information for the observation of a hot star.

\subsection{Visibility}

``Visibility'', i.e. the
``normalised-amplitude-of-the-Fourier-transform-of-the-space-distribution-of-light-of-an-observed-object''
in its common accepted definition, is a number between zero and
one. The bigger the object is at a given baseline and wavelength, the
lower the visibility is. In the same way, for a given object and
wavelength, the bigger the baseline is, the lower the visibility will
be. That is why it is often used as a probe of the typical angular
size of an object.

On the other hand, the behaviour along wavelength will be somewhat
more complicated. As an example, H-band is commonly more sensitive to
stellar photosphere, whereas N-band will probe more dusty structures,
which are commonly much larger than the stars. The visibility
may increase, or decrease with larger wavelength, whether there is a
dusty envelope around a star or not. Therefore, the
wavelength-dependent behaviour of the visibility is not
straightforward and will highly depend on the nature of the object
observed.

\subsection{Phases}

Absolute phase measurements with optical interferometry are either
very difficult or impossible to achieve due to the Earth's atmosphere
blurring. Therefore, very often, the only available phase measurements
are partial, through the closure phase and/or the differential phase.

\subsection{Closure phase}

Closure phase is a phase measurement made available when combining
three telescopes in a triangle. Its main advantage is that it is
mostly insensitive to any atmospheric effect. Its main drawback is
that it measure 1/3 of the whole phase information out of the three
baselines available. It is commonly accepted as a probe for
``asymmetry'' of the object when it is non-zero. However an object can
be asymmetric and exhibit a zero closure phase for a given baseline
and a given wavelength. Hence a zero closure phase is not always a
sign of the object being symmetric. As a summary, closure phase is a 
tool to probe for asymmetric structures, such as inhomogeneities in
circumstellar disks \citet{2007A&A...464...73M}, or in the stellar
photosphere itself \citep{2008A&A...489L...5D}.

\subsection{Differential phase}

Differential phase is the wavelength-dependent variation of the phase
that can be partially reconstructed out of the interferometric
measurements, as soon as there is a spectrograph inhere. It provide
invaluable phase information as it still contains the
wavelength-dependency of the phases in e.g. an emission line
\citep{2007A&A...464...59M}, at the cost of loosing the absolute
(average) level of the phase on the whole wavelength-range
considered. Hence, it is possible to probe the change of asymmetry of
an object as a function of wavelength with such a measure
\citep{2007A&A...464..107M}.

\section{What is the angular size of a massive hot star?}
\label{sec:}

The question how to measure the size of stars has been raised by
\citet{Fizeau1851}, which induced a series of unsuccessful
observations, made by \citet{Stephan1874}, and led to an upper
limit on the size of stars. It has been only when larger instruments
became available that the first star, Betelgeuse, could be resolved by
\citet{1921ApJ....53..249M}. However, until the reborn of
interferometry by \citet{1975ApJ...196L..71L}, only very few stars
could have their angular diameters measured. The question whether one
can measure the angular size of hot stars with the modern technique of
interferometry covers in fact two issues, which are to resolve the
stellar photosphere itself, and to resolve the circumstellar envelope
or wind, which are sometimes the dominant source of emission at some
specific wavelengths.

\subsection{B-type stars}

B-type massive stars have usually a very small angular diameter. This
is due to the fact that main-sequence stars have small physical sizes,
but can be found close to us, and that giant or supergiant stars are
bigger in physical sizes, but are sparser in the Galaxy and hence far
away from Earth. An example of such is the closest Be star, $\alpha$
Arae, which has a radius of 4.8\rsun{} and is located at 74\,pc from
the Earth. Converted in angular diameter, this gives a diameter of
0.6\,mas, which is slightly above the capacities in resolution for the
VLTI. On the other hand, the disk-like envelope of this star was
measured by AMBER to be $\simeq$4\,mas in radius, which is well within
the possibilities of the VLTI.

A (highly non-exhaustive) list of Be stars and B[e] supergiant stars,
observed up to now, is presented in table \ref{tab:BstarsObserved},
and shows that for the brightest and closest high mass B-type stars,
the stellar photosphere itself is always smaller than 1\,mas. On the
other hand, the envelope sizes of these stars range from less than
1\,mas for the smallest one up to $\approx$10\,mas for the largest
ones, making possible to image some of them with the current
capabilities of the interferometers. This also explains why most of
the interferometric studies on massive B-type stars have focused on
their circumstellar envelopes and not really on their stellar disk
itself.

\begin{table*}[htbp]\centering
  \setlength{\tabnotewidth}{\columnwidth}
  \tablecols{3}
  \setlength{\tabcolsep}{2.8\tabcolsep}
  \caption[Interferometers sensitivity]{
    \footnotesize{
      A non-exhaustive list of hot stars observed with interferometry
      (in near-infrared) and their measured [or estimated] sizes.
    }
  }
  \begin{tabular}{|c|c|c|c|}
    \hline
    Name & Star \o& Envelope \o & Reference \\
     & (mas)& (mas)&\\
    \hline
    \hline
    \multicolumn{4}{|l|}{\bf B-type stars} \\
    \hline
    $\kappa$ CMa      & [0.24]  & 11   & \protect{\citet{2007A&A...464...73M}}  \\
    MWC 297           & [0.22]  & 10   & \protect{\citet{2007A&A...464...43M}}  \\
    $\alpha$ Arae     & [0.6]   & 8    & \protect{\citet{2007A&A...464...59M}}  \\
    HD87643           & -       & 4    & \protect{\citet{MillourInpreparation.}}  \\
    CPD-52$^\circ$2874 & -       & 3.4  & \protect{\citet{2007A&A...464...81D}}  \\
    $\zeta$ Tau       & 0.4-[0.5]     & 2 &
    \protect{\citet{2007ApJ...654..527G, 2009arXiv0901.1098C}}  \\
    $\kappa$ Dra      & 0.39    & 1.72 & \protect{\citet{2007ApJ...654..527G}}  \\
    $\delta$ Cen      & [0.45]  & 1.6  & \protect{\citet{2008A&A...488L..67M}}  \\
    $\gamma$ Cas      & 0.5     & 1.36 & \protect{\citet{2007ApJ...654..527G}}  \\
    $\phi$ Per        & 0.3     & 0.51 & \protect{\citet{2007ApJ...654..527G}}  \\
    \hline
    \hline
    \multicolumn{4}{|l|}{\bf O-type stars} \\
    \hline
    Rigel ($\beta$ Ori)      & [2.43]  & - & \protect{\citet{2005A&A...431..773R}}  \\
    O star of $\gamma^2$ Vel & [0.48] &-&  \protect{\citet{2007A&A...464..107M}}  \\
    $\theta^1$ Ori C         & [0.22] &-& \protect{\citet{2007A&A...474..515M, 2006A&A...448..351S}}  \\
    \hline
    \hline
    \multicolumn{4}{|l|}{\bf WRs, LBVs, etc.} \\
    \hline
    WR star of $\gamma$ Vel &- & [0.3-0.6] & \protect{\citet{2007A&A...464..107M}}  \\
    $\eta$ Car      &-  & 2.2-3.8 & \protect{\citet{2007A&A...464...87W}}  \\
    \hline
  \end{tabular}
      \label{tab:BstarsObserved}
\end{table*}

\begin{figure*}[htbp]
  \includegraphics[width=\columnwidth]{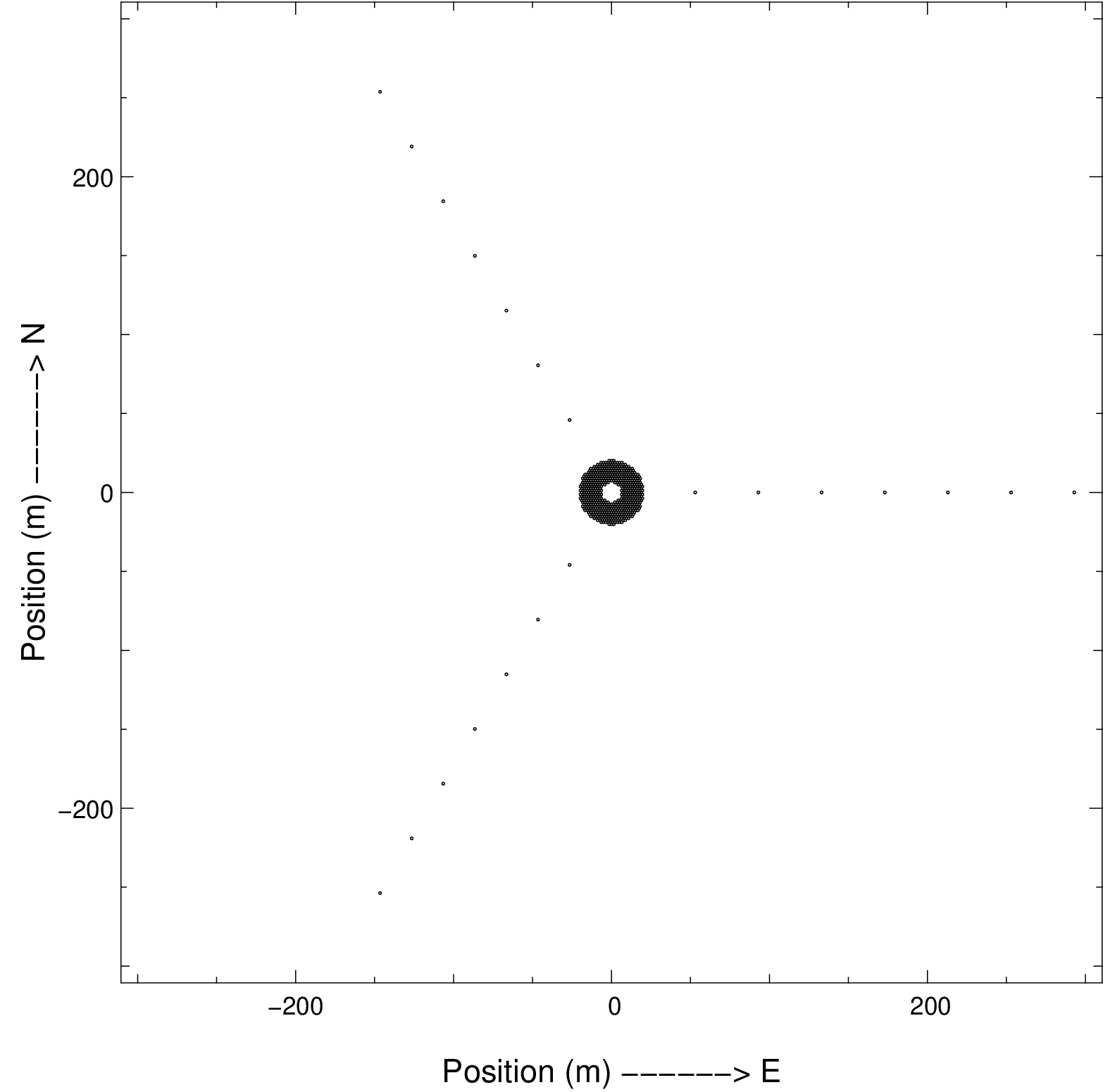}%
  \hspace*{\columnsep}%
  \includegraphics[width=\columnwidth]{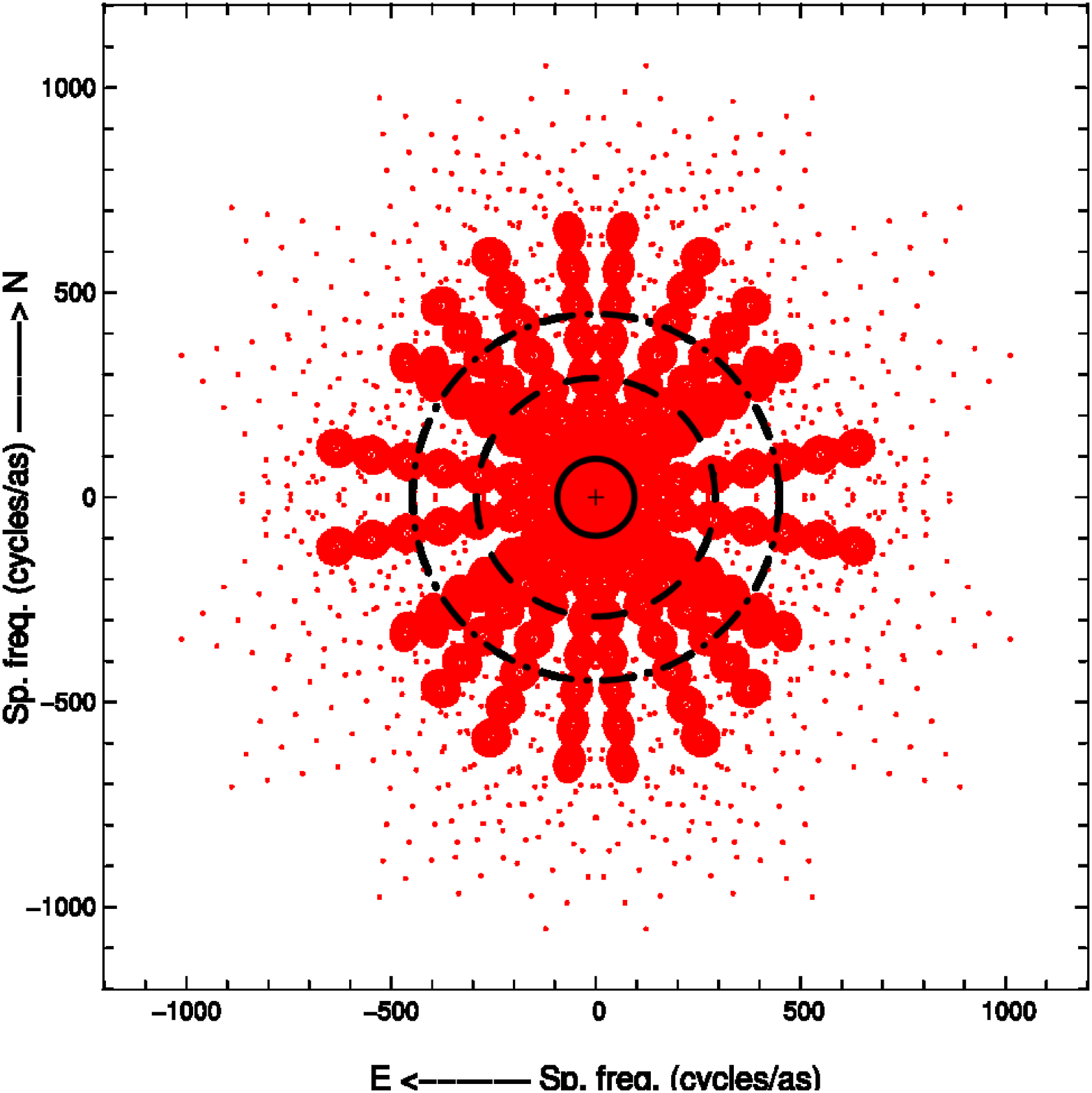}
  \caption{
    A sketch of the perfect interferometer: it would combine the
    extreme collecting area of the E-ELT with large
    baselines, with a set of fixed auxilliary telescopes (ATs), made
    e.g. using the same segments as in the E-ELT. High fidelity images
    at an angular resolution of 0.45\,mas in K band may be
    reconstructed in this configuration with only 3 different hour
    angles ($\approx$3\,h of observing time, see UV map on the
    right). The center circle show the UV frequencies probed by the
    E-ELT alone, the dashed circle show the same for the VLTI (130\,m
    maximum baseline) and the dash-dotted circle represent the maximum
    reachable baseline of the VLTI (200\,m), with an upgrade of its
    delay lines.
  }
  \label{fig:E-ELTi}
\end{figure*}

\subsection{O-type stars}

Among the regions where one can find massive O-type stars, the Orion
nebula is maybe one of the closest places, with a distance of
$414\pm7$\,pc \citep{2007A&A...474..515M}. The young B- and O-type
stars of the Orion trapezium, and especially $\theta^1$~Ori~C, are
at the origin of the ionization of the famous nebula itself. If one
take the expected radius for such a star \citep[$R_{*} = 9.9
R_{\odot}$][]{2006A&A...448..351S}, this gives an angular diameter of
0.22\,mas. A similar estimate (0.15\,mas) can also be found in the
CADARS catalog of stellar angular diameters
\citep{2001A&A...367..521P}. Other O-type stars are listed in table
\ref{tab:BstarsObserved} showing that only Rigel ($\beta$ Ori),
which has an angular diameter of 2.43\,mas
\citep{2005A&A...431..773R}, can be resolved with current
facilities. This means that very long baselines combined with shorter
wavelengths are needed to resolve these stars or their close-by
co-rotating wind features (only a few $R_{*}$).

\subsection{Wolf-Rayet and other exotic-types stars}

Wolf-Rayet stars are evolved massive stars, thought to be progenitors
of supernovae. They are among the hottest stars with temperatures
$\approx$70,000\,K and no direct information on the underlying star
can be obtained with spectroscopy as its fast
($v_\infty\approx1500-2000$\,km$\times$s$^{-1}$) and dense wind
completely veils the star. Therefore, fundamental parameters such as
their luminosity and mass can be accessed through complex radiative
transfer models, extensively tested with spectroscopy, but that still
need to be checked by direct measurements. Interferometry allows to
resolve the line forming regions and probe the models. Also, some WR
stars exhibit a large and dusty circumstellar shell that can be probed
with interferometry \citep{2009Msngr.135...26M}. Angular diameters of
WR stars and their wind are very small (see
Table~\ref{tab:BstarsObserved}), therefore, only circumstellar
features are today reachable with interferometry. For LBV stars, the
situation is slightly better since their wind is denser and, hence,
appear larger.

\section{Bigger and larger interferometers for the future?}
\label{sec:}

The today's interferometric facilities have fairly bright limiting
magnitudes compared to single dish telescopes. Both the VLTI and
Keck-I have extreme limiting magnitudes lying in the 10-11\,mag range
in the K-band (Kishimoto, Weigelt, private communication). At these
limits, expected relative uncertainties will be larger than 10\%,
which would not be suitable to resolve fainter (and hence smaller) hot
stars. More conservative limiting magnitudes ($K=7$ for VLTI with UTs,
$K=5$ for VLTI with ATs) are usually adopted. One has to note that the
V band limit has also to be taken into account ($V\leq13.5$ for the
VLTI). As an illustration, the reachable number of Wolf-Rayet stars
with the offered VLTI performances is about 10.

Another limitation of current facilities is the fact that current
baselines are limited to few hundred meters, still making stellar
surfaces resolution above the current capacities, as explained in the
previous sections. One cannot combine today at the same time large
baselines (like in the CHARA array) and large collecting areas (like
in the VLTI and the Keck-I).

Having limiting magnitudes fainter than e.g. 15 in K combined with
baselines larger than e.g. 300\,m would increase the number of
potential targets for a given object class from few dozens to many
thousands, allowing one to access the statistical aspect of hot stars
astrophysics. This might be reached with a combination of
higher-throughput facilities and instruments, the generalization of
dual field interferometry, and of course larger collectig areas and
larger baselines lengths. Sketch of such future facilities have been
already described \citep{1988ESOC...29..669L, 2001LIACo..36...85R,
  2003sf2a.conf..365V}, but they all combine relatively small
apertures (D$\approx$2\,m). Accessing fainter magnitudes might be
reached by combining ELTs with interferometry. This might be done via
a set of fixed auxilliary telescopes dispatched around an ELT. A
sketch of such is shown in Fig.~\ref{fig:E-ELTi}.

%

\section*{\bf Q. \& A. during the conference:}
\begin{enumerate}
\item[E. Trunkowski:] How  many stars were oberved with
  interferometers up to now?
\item[F. Millour:] This largely depends on the type of objects
  considered. In total, few thousand stars are reachable with
  current facilities. $\approx$1500 targets were observed
  with the VLTI. The raw data can be accessed through the ESO
  archive\footnote{\url{http://archive.eso.org}} for all stars, after
  a proprietary time of one year. There is also a PTI and Keck-I
    archive\footnote{\url{http://nexsci.caltech.edu}}.
\item[D. Baade:] What future for VLTI in the E-ELT era?
\item[M. Schoeller:] The large difference in collecting
  area (VLTI$\ll$E-ELT) as well as angular resolution (VLTI$\gg$E-ELT)
  make these two facilities complementary.
\end{enumerate}


\begin{thebibliography}

\bibitem[{{Carciofi} {et~al.}(2009){Carciofi}, {Okazaki}, {le Bouquin}, {{\v
  S}tefl}, {Rivinius}, {Baade}, {Bjorkman}, \& {Hummel}}]{2009arXiv0901.1098C}
{Carciofi}, A.~C., {et~al.} 2009,
  A\&A, In press.


\bibitem[{{Domiciano de Souza} {et~al.}(2008){Domiciano de Souza}, {Bendjoya},
  {Vakili}, {Millour}, \& {Petrov}}]{2008A&A...489L...5D}
{Domiciano de Souza}, A., {et~al.} 2008, \aap, 489, L5


\bibitem[{{Domiciano de Souza} {et~al.}(2007){Domiciano de Souza}, {Driebe},
  {Chesneau}, {Hofmann}, {Kraus}, {Miroshnichenko}, {Ohnaka}, {Petrov},
  {Preisbisch}, {Stee}, {Weigelt}, {Lisi}, {Malbet}, \&
  {Richichi}}]{2007A&A...464...81D}
{Domiciano de Souza}, A. {et~al.} 2007, \aap, 464, 81


\bibitem[{{Fizeau}(1851)}]{Fizeau1851}
{Fizeau}, H. 1851, Sur un moyen de d\'eduire les diam\`etres des \'etoiles de
  certains ph\'enom\`enes d'interf\'erence


\bibitem[{{Gies} {et~al.}(2007){Gies}, {Bagnuolo}, {Baines}, {ten Brummelaar},
  {Farrington}, {Goldfinger}, {Grundstrom}, {Huang}, {McAlister}, {M{\'e}rand},
  {Sturmann}, {Sturmann}, {Touhami}, {Turner}, {Wingert}, {Berger}, {McSwain},
  {Aufdenberg}, {Ridgway}, {Cochran}, {Lester}, {Sterling}, {Bjorkman},
  {Bjorkman}, \& {Koubsk{\'y}}}]{2007ApJ...654..527G}
{Gies}, D.~R., {et~al.}  2007, \apj, 654, 527


\bibitem[{{Haniff}(2007)}]{2007NewAR..51..565H}
{Haniff}, C. 2007, New Astronomy Review, 51, 565


\bibitem[{{Labeyrie}(1975)}]{1975ApJ...196L..71L}
{Labeyrie}, A. 1975, \apjl, 196, L71


\bibitem[{{Labeyrie} {et~al.}(1988){Labeyrie}, {Lemaitre}, {Thom}, \&
  {Vakili}}]{1988ESOC...29..669L}
{Labeyrie}, A., {et~al.} 1988, in ESO Astrophysics Symposia, 29,
669--693

\bibitem[{{Malbet} {et~al.}(2007){Malbet}, {Benisty}, {de Wit}, {Kraus},
  {Meilland}, {Millour}, {Tatulli}, {Berger}, {Chesneau}, {Hofmann}, {Isella},
  {Natta}, {Petrov}, {Preibisch}, {Stee}, {Testi}, {Weigelt}, {Antonelli},
  {Beckmann}, {Bresson}, {Chelli}, {Dugu{\'e}}, {Duvert}, {Gennari},
  {Gl{\"u}ck}, {Kern}, {Lagarde}, {Le Coarer}, {Lisi}, {Perraut}, {Puget},
  {Rantakyr{\"o}}, {Robbe-Dubois}, {Roussel}, {Zins}, {Accardo}, {Acke},
  {Agabi}, {Altariba}, {Arezki}, {Aristidi}, {Baffa}, {Behrend}, {Bl{\"o}cker},
  {Bonhomme}, {Busoni}, {Cassaing}, {Clausse}, {Colin}, {Connot},
  {Delboulb{\'e}}, {Domiciano de Souza}, {Driebe}, {Feautrier}, {Ferruzzi},
  {Forveille}, {Fossat}, {Foy}, {Fraix-Burnet}, {Gallardo}, {Giani}, {Gil},
  {Glentzlin}, {Heiden}, {Heininger}, {Hernandez Utrera}, {Kamm}, {Kiekebusch},
  {Le Contel}, {Le Contel}, {Lesourd}, {Lopez}, {Lopez}, {Magnard}, {Marconi},
  {Mars}, {Martinot-Lagarde}, {Mathias}, {M{\`e}ge}, {Monin}, {Mouillet},
  {Mourard}, {Nussbaum}, {Ohnaka}, {Pacheco}, {Perrier}, {Rabbia}, {Rebattu},
  {Reynaud}, {Richichi}, {Robini}, {Sacchettini}, {Schertl}, {Sch{\"o}ller},
  {Solscheid}, {Spang}, {Stefanini}, {Tallon}, {Tallon-Bosc}, {Tasso},
  {Vakili}, {von der L{\"u}he}, {Valtier}, {Vannier}, \&
  {Ventura}}]{2007A&A...464...43M}
{Malbet}, F., {et~al.}
  2007, \aap, 464, 43


\bibitem[{{Meilland} {et~al.}(2007{\natexlab{a}}){Meilland}, {Millour}, {Stee},
  {Domiciano de Souza}, {Petrov}, {Mourard}, {Jankov}, {Robbe-Dubois}, {Spang},
  {Aristidi}, {Antonelli}, {Beckmann}, {Bresson}, {Chelli}, {Dugu{\'e}},
  {Duvert}, {Gennari}, {Gl{\"u}ck}, {Kern}, {Lagarde}, {Le Coarer}, {Lisi},
  {Malbet}, {Perraut}, {Puget}, {Rantakyr{\"o}}, {Roussel}, {Tatulli},
  {Weigelt}, {Zins}, {Accardo}, {Acke}, {Agabi}, {Altariba}, {Arezki}, {Baffa},
  {Behrend}, {Bl{\"o}cker}, {Bonhomme}, {Busoni}, {Cassaing}, {Clausse},
  {Colin}, {Connot}, {Delboulb{\'e}}, {Driebe}, {Feautrier}, {Ferruzzi},
  {Forveille}, {Fossat}, {Foy}, {Fraix-Burnet}, {Gallardo}, {Giani}, {Gil},
  {Glentzlin}, {Heiden}, {Heininger}, {Hernandez Utrera}, {Hofmann}, {Kamm},
  {Kiekebusch}, {Kraus}, {Le Contel}, {Le Contel}, {Lesourd}, {Lopez}, {Lopez},
  {Magnard}, {Marconi}, {Mars}, {Martinot-Lagarde}, {Mathias}, {M{\`e}ge},
  {Monin}, {Mouillet}, {Nussbaum}, {Ohnaka}, {Pacheco}, {Perrier}, {Rabbia},
  {Rebattu}, {Reynaud}, {Richichi}, {Robini}, {Sacchettini}, {Schertl},
  {Sch{\"o}ller}, {Solscheid}, {Stefanini}, {Tallon}, {Tallon-Bosc}, {Tasso},
  {Testi}, {Vakili}, {von der L{\"u}he}, {Valtier}, {Vannier}, \&
  {Ventura}}]{2007A&A...464...73M}
{Meilland}, A., {et~al.} 
  2007{\natexlab{a}}, \aap, 464, 73


\bibitem[{{Meilland} {et~al.}(2008){Meilland}, {Millour}, {Stee}, {Spang},
  {Petrov}, {Bonneau}, {Perraut}, \& {Massi}}]{2008A&A...488L..67M}
{Meilland}, A., {et~al.} 2008, \aap, 488, L67


\bibitem[{{Meilland} {et~al.}(2007{\natexlab{b}}){Meilland}, {Stee}, {Vannier},
  {Millour}, {Domiciano de Souza}, {Malbet}, {Martayan}, {Paresce}, {Petrov},
  {Richichi}, \& {Spang}}]{2007A&A...464...59M}
{Meilland}, A., {et~al.} 2007{\natexlab{b}}, \aap, 464, 59


\bibitem[{{Menten} {et~al.}(2007){Menten}, {Reid}, {Forbrich}, \&
  {Brunthaler}}]{2007A&A...474..515M}
{Menten}, K.~M., {et~al.} 2007, \aap,
  474, 515


\bibitem[{{Michelson} \& {Pease}(1921)}]{1921ApJ....53..249M}
{Michelson}, A.~A. \& {Pease}, F.~G. 1921, \apj, 53, 249


\bibitem[{{Millour} {et~al.} (In preparation.)}]{MillourInpreparation.}
{Millour} {et~al.} In preparation., A\&A


\bibitem[{{Millour}(2008)}]{2008NewAR..52..177M}
{Millour}, F. 2008, New Astronomy Review, 52, 177


\bibitem[{{Millour} {et~al.}(2009){Millour}, {Chesneau}, {Driebe}, {Matter},
  {Schmutz}, {Lopez}, {Petrov}, {Groh}, {Bonneau}, {Dessart}, {Hofmann}, \&
  {Weigelt}}]{2009Msngr.135...26M}
{Millour}, F., {et~al.} 2009, The Messenger, 135, 26


\bibitem[{{Millour} {et~al.}(2007){Millour}, {Petrov}, {Chesneau}, {Bonneau},
  {Dessart}, {Bechet}, {Tallon-Bosc}, {Tallon}, {Thi{\'e}baut}, {Vakili},
  {Malbet}, {Mourard}, {Antonelli}, {Beckmann}, {Bresson}, {Chelli},
  {Dugu{\'e}}, {Duvert}, {Gennari}, {Gl{\"u}ck}, {Kern}, {Lagarde}, {Le
  Coarer}, {Lisi}, {Perraut}, {Puget}, {Rantakyr{\"o}}, {Robbe-Dubois},
  {Roussel}, {Tatulli}, {Weigelt}, {Zins}, {Accardo}, {Acke}, {Agabi},
  {Altariba}, {Arezki}, {Aristidi}, {Baffa}, {Behrend}, {Bl{\"o}cker},
  {Bonhomme}, {Busoni}, {Cassaing}, {Clausse}, {Colin}, {Connot},
  {Delboulb{\'e}}, {Domiciano de Souza}, {Driebe}, {Feautrier}, {Ferruzzi},
  {Forveille}, {Fossat}, {Foy}, {Fraix-Burnet}, {Gallardo}, {Giani}, {Gil},
  {Glentzlin}, {Heiden}, {Heininger}, {Hernandez Utrera}, {Hofmann}, {Kamm},
  {Kiekebusch}, {Kraus}, {Le Contel}, {Le Contel}, {Lesourd}, {Lopez}, {Lopez},
  {Magnard}, {Marconi}, {Mars}, {Martinot-Lagarde}, {Mathias}, {M{\`e}ge},
  {Monin}, {Mouillet}, {Nussbaum}, {Ohnaka}, {Pacheco}, {Perrier}, {Rabbia},
  {Rebattu}, {Reynaud}, {Richichi}, {Robini}, {Sacchettini}, {Schertl},
  {Sch{\"o}ller}, {Solscheid}, {Spang}, {Stee}, {Stefanini}, {Tasso}, {Testi},
  {von der L{\"u}he}, {Valtier}, {Vannier}, \& {Ventura}}]{2007A&A...464..107M}
{Millour}, F., {et~al.} 2007, \aap, 464, 107


\bibitem[{{Pasinetti Fracassini} {et~al.}(2001){Pasinetti Fracassini},
  {Pastori}, {Covino}, \& {Pozzi}}]{2001A&A...367..521P}
{Pasinetti Fracassini}, L.~E., {et~al.}
  2001, \aap, 367, 521


\bibitem[{{Riaud} {et~al.}(2001){Riaud}, {Gillet}, {Labeyrie}, {Boccaletti},
  {Schneider}, {Rouan}, {Baudrand}, {Arnold}, {Borkowski}, \&
  {Lardi{\`e}re}}]{2001LIACo..36...85R}
{Riaud}, P., {et~al.} 2001, Liege International Astrophysical Colloquia,
85--95


\bibitem[{{Richichi} {et~al.}(2005){Richichi}, {Percheron}, \&
  {Khristoforova}}]{2005A&A...431..773R}
{Richichi}, A., {et~al.}  2005, \aap, 431, 773


\bibitem[{{Sim{\'o}n-D{\'{\i}}az} {et~al.}(2006){Sim{\'o}n-D{\'{\i}}az},
  {Herrero}, {Esteban}, \& {Najarro}}]{2006A&A...448..351S}
{Sim{\'o}n-D{\'{\i}}az}, S.,  {et~al.}
  2006, \aap, 448, 351


\bibitem[{{Stephan}(1874)}]{Stephan1874}
{Stephan}, E. 1874, in C. R. Acad. Sci., Vol.~78, 1008


\bibitem[{{Vakili} {et~al.}(2003){Vakili}, {Aristidi}, {Fossat}, {Abe},
  {Domiciano}, {Belu}, {Agabi}, {Schmider}, {Lopez}, \&
  {Swain}}]{2003sf2a.conf..365V}
{Vakili}, F., {et~al.} 2003, SF2A, 365

\bibitem[{{Weigelt} {et~al.}(2007){Weigelt}, {Kraus}, {Driebe}, {Petrov},
  {Hofmann}, {Millour}, {Chesneau}, {Schertl}, {Malbet}, {Hillier}, {Gull},
  {Davidson}, {Domiciano de Souza}, {Antonelli}, {Beckmann}, {Bresson},
  {Chelli}, {Dugu{\'e}}, {Duvert}, {Gennari}, {Gl{\"u}ck}, {Kern}, {Lagarde},
  {Le Coarer}, {Lisi}, {Perraut}, {Puget}, {Rantakyr{\"o}}, {Robbe-Dubois},
  {Roussel}, {Tatulli}, {Zins}, {Accardo}, {Acke}, {Agabi}, {Altariba},
  {Arezki}, {Aristidi}, {Baffa}, {Behrend}, {Bl{\"o}cker}, {Bonhomme},
  {Busoni}, {Cassaing}, {Clausse}, {Colin}, {Connot}, {Delboulb{\'e}},
  {Feautrier}, {Ferruzzi}, {Forveille}, {Fossat}, {Foy}, {Fraix-Burnet},
  {Gallardo}, {Giani}, {Gil}, {Glentzlin}, {Heiden}, {Heininger}, {Hernandez
  Utrera}, {Kamm}, {Kiekebusch}, {Le Contel}, {Le Contel}, {Lesourd}, {Lopez},
  {Lopez}, {Magnard}, {Marconi}, {Mars}, {Martinot-Lagarde}, {Mathias},
  {M{\`e}ge}, {Monin}, {Mouillet}, {Mourard}, {Nussbaum}, {Ohnaka}, {Pacheco},
  {Perrier}, {Rabbia}, {Rebattu}, {Reynaud}, {Richichi}, {Robini},
  {Sacchettini}, {Sch{\"o}ller}, {Solscheid}, {Spang}, {Stee}, {Stefanini},
  {Tallon}, {Tallon-Bosc}, {Tasso}, {Testi}, {Vakili}, {von der L{\"u}he},
  {Valtier}, {Vannier}, {Ventura}, {Weis}, \&
  {Wittkowski}}]{2007A&A...464...87W}
{Weigelt}, G.,  2007, {et~al.}
  \aap, 464, 87



\end{thebibliography}
\end{document}